\documentclass[aps,twocolumn,showpacs]{revtex4}
\usepackage{graphicx}
\usepackage{dcolumn}
\usepackage{bm}
\usepackage{amsmath}
\begin{document}

\title {Numerical Study of Transition to Stripe Phases
in High Temperature Superconductors Under a Strong Magnetic Field}
\author{Yan Chen, H. Y. Chen and C. S. Ting}
\affiliation{Texas Center for Superconductivity and Department of
Physics, University of Houston, Houston, TX 77204}

\begin{abstract}
{The nature of spin density wave (SDW) and charge density wave
(CDW) in the mixed state of high $T_c$ superconductors (HTS) is
investigated by using the self-consistent Bogoliubov-de Gennes
equations and an effective model Hamiltonian with competing SDW
and $d$-wave superconductivity interactions. We show that there
exists a critical onsite Coulomb interaction $U_c$. For optimally
doped sample, a two dimensional SDW and CDW modulations are
induced for $U < U_c$ while SDW and CDW orders become
antiferromagnetic (AF) and charge stripes for $U
> U_c$. These stripe orders are stabilized and enhanced near the
vortex cores. The wavelengths of the AF stripes and charge stripes
are found respectively to be $8 a$ and $4 a$ with $a$ as the
lattice constant. We show that our results could be applied to
understand several recent experiments on HTS.}
\end{abstract}


\maketitle

Intensive efforts have been focused on the searching and the
understanding of spin-density wave (SDW) and other phases in the
mixed state of high $T_c$ superconductors (HTS) for the past
several years.  Experiments from the neutron
scattering~\cite{Katano,Lake01,Lake02}, scanning tunneling
microscope(STM)~\cite{Maggio,Pan00,Hoff1},  and nuclear magnetic
resonance(NMR)~\cite{NMR1} provided vital information on these
topics. For example, according to the neutron scattering
experiment by Lake {\em et al.}~\cite{Lake01}, a remarkable
antiferromagnetism or SDW appears in the optimally doped
La$_{2-x}$Sr$_x$CuO$_4$ when a strong magnetic field is applied.
Recently, Hoffman {\em et al.}~\cite{Hoff1} studied the LDOS in
the mixed states of optimally doped Bi$_2$Sr$_2$CaCu$_2$O$_{8+
\delta}$ using STM measurements, and they found that associated
with the SDW, anisotropic charge density wave (CDW) exists both
inside and outside the vortex cores. The coexistence of $d$-wave
superconductivity (DSC), SDW and CDW orders in terms of the stripe
phases was theoretically studied in the absence of a magnetic
field ~\cite{Emery,Bale98,Sach99,Mart00}.

Although the competition between SDW and DSC in a magnetic field
was previously examined~\cite{Arovas97,Machida,Demler,Lee01}, the
nature of the SDW and CDW and their spatial variations have not
been addressed in such detail as to compare with the experiments.
In this paper, we shall adopt the method described in previous
papers~\cite{Wang95,Zhu01} to examine the possible existence of
SDW and accompanying CDW orders in the mixed state of HTS, and
their nature in optimally doped samples. In order to simplify the
numerical calculation, we shall assume a square vortex lattice for
the mixed state and a strong magnetic field $B$ such that $
\lambda \gg b \gg \xi$ with $\lambda$ as the London penetration
depth, $\xi$ the coherence length and $b$ the vortex lattice
constant. Under this condition, the applied magnetic field $B$ can
be regarded as a constant throughout the sample. Our calculation
is based upon a model Hamiltonian with competing DSC and SDW
orders and realistic band structure parameters. For an optimally
doped sample $(x=0.15)$ under a strong magnetic field $B$, we find
that DSC, SDW and CDW stripe phases could be in existence and they
are pinned and enhanced  by the vortex lattice. Numerical
calculation based upon a magnetic unit cell of $48 \times 24$
lattice sites shows the wavelength of the SDW stripe to be $8 a$
and the accompanying CDW stripe to be $4 a$. A phenomenological
model will be used to explain the anisotropic SDW and CDW observed
by experiments~\cite{Lake01,Hoff1}.

Let us begin with an effective mean field model in which
interactions describing both DSC and antiferromagnetic (AF) orders
in a two-dimensional square lattice are considered. The effective
one band Hamiltonian can be written as:
\begin{eqnarray}
H&=&\sum_{{\bf i,j},\sigma} - {t_{\bf i,j}}  c_{{\bf
i}\sigma}^{\dagger}c_{{\bf j}\sigma} +\sum_{{\bf i},\sigma}( U
n_{{\bf i} {\bar {\sigma}}} -\mu)c_{{\bf i}\sigma}^{\dagger}
c_{{\bf i}\sigma} \nonumber \\ &&+\sum_{\bf i,j} ( {\Delta_{\bf
i,j}} c_{{\bf i}\uparrow}^{\dagger} c_{{\bf
j}\downarrow}^{\dagger} + h.c.)\;.
\end{eqnarray}
where $c_{{\bf i}\sigma}^{\dagger}$ is the electron creation
operator and $\mu$ is the chemical potential. In the presence of
magnetic field B, the hopping integral can be expressed as $
t_{\bf i,j}= t_{\bf i,j}^{0} exp[{i \frac{\pi}{\Phi_{0}}
\int_{{\bf r}_{\bf j}}^{{\bf r}_{\bf i}} {\bf A}({\bf r})\cdot
d{\bf r}}]$ where $t_{\bf i,j}^{0} = t$ for the nearest
neighboring sites $(i,j)$ while the next-nearest neighbor hopping
$t_{\bf i,j}^{0} = t^{\prime}$. The superconducting flux quanta
denotes as $\Phi_0=h/2e$. Here we choose Landau gauge ${\bf
A}=(-By,0,0)$ with $y$ as the $y$-component of the position vector
{\bf r}. The two possible orders in cuprates are SDW and DSC which
have the following definitions respectively: $\Delta^{SDW}_{\bf i}
= U \langle c_{{\bf i} \uparrow}^{\dagger} c_{{\bf i} \uparrow}
-c_{{\bf i} \downarrow}^{\dagger}c_{{\bf i} \downarrow} \rangle$
and $\Delta_{\bf i,j}=V_{DSC} \langle c_{{\bf i}\uparrow}c_{{\bf
j}\downarrow}-c_{{\bf i} \downarrow} c_{{\bf j}\uparrow} \rangle
/2$. In the above expressions, $U$ and $V_{DSC}$ are respectively
the interaction strengths for SDW and DSC orders. $V_{DSC}$, which
gives rise to the d-wave superconductivity, may come from all
possibilities including AF fluctuations and electron phonon
interactions. The mean-field Hamiltonian (1) can be diagonalized
by solving the resulting Bogoliubov-de Gennes (BdG) equations
self-consistently
\begin{equation}
\sum_{\bf j} \left(\begin{array}{cc} {\cal H}_{\bf i,j}&
\Delta_{\bf i,j} \\
\Delta_{\bf i,j}^{*} & -{\cal H}_{\bf i,j}^{*}
\end{array}
\right) \left(\begin{array}{c} u_{\bf j}^{n} \\ v_{\bf j}^{n}
\end{array}
\right) =E_{n} \left(
\begin{array}{c}
u_{\bf i}^{n} \\ v_{\bf i}^{n}
\end{array}
\right)\;,
\end{equation}
where the single particle Hamiltonian ${\cal H}_{\bf
i,j}^{\sigma}= -t_{\bf i,j} +(U n_{{\bf i} \bar{\sigma}}
-\mu)\delta_{\bf ij}$, and and $n_{i \uparrow} = \sum_{n}
|u_{i\uparrow}^{n}|^2 f(E_{n})$, $ n_{i \downarrow} = \sum_{n}
|v_{i\downarrow}^{n}|^2 ( 1- f(E_{n}))$, $ \Delta_{ij} =
\frac{V_{DSC}} {4} \sum_{n} (u_{i\uparrow}^{n}
v_{j\downarrow}^{n*} +v_{i\downarrow}^{*} u_{j\uparrow}^{n}) \tanh
\left( \frac{E_{n}} {2k_{B}T} \right)$, with $f(E)$ as the Fermi
distribution function and the electron density $n_{\bf i}= n_{{\bf
i} \uparrow} + n_{{\bf i} \downarrow}$. The DSC order parameter is
defined at site $i$ as $\Delta^{D}_{\bf i}= (\Delta^{D}_{\bf
i+e_x,i} + \Delta^{D}_{\bf i-e_x,i} - \Delta^{D}_{\bf i,i+e_y}
-\Delta^{D}_{\bf i,i-e_y})/4$ where $ \Delta^{D}_{\bf i,j} =
\Delta_{\bf i,j} exp[ i { \frac{\pi}{\Phi_{0}} \int_{{\bf r}_{\bf
i}}^{({\bf r}_{\bf i}+{\bf r}_{\bf j})/2 } {\bf A}({\bf r}) \cdot
d{\bf r}}]$ and ${\bf e}_{x,y}$ denotes the unit vector along
$(x,y)$ direction. The main procedure of self-consistent
calculation is given below: For a given initial set of parameters
$n_{{\bf i} \sigma}$ and $\Delta_{\bf i, j}$, the Hamiltonian is
numerically diagonalized and the electron wave functions obtained
are used to calculate the new parameters for the next iteration
step. The calculation is repeated until the relative difference of
order parameter between two consecutive iteration step is less
than $10^{-4}$. By varying the chemical potential, one obtains
solutions corresponding to various doping concentrations.

In the following calculation, the length and energy are measured
in units of the lattice constant $a$ and the hopping integral $t$
respectively. Here the next-nearest neighboring hopping integral
is chosen to be $t^{\prime}= -0.2$ to fit the band structure of
HTS. It needs to be pointed out that the induction of internal
magnetic field by the supercurrent around the vortex core is very
small as compared with the external magnetic field, so that the
uniform magnetic field distribution is a valid approximation. We
follow the standard procedures~\cite{Wang95,Zhu01} to introduce
magnetic unit cells, where each unit cell accommodates two
superconducting flux quanta. Periodic boundary condition is
imposed in calculation.  The related parameters are chosen as the
following: For optimal doping $x = 0.15$ (or the electron doping
$n_f = 0.85$), the DSC coupling strength is $V_{DSC}=1.0$, the
linear dimension of the unit cell of the vortex lattice is chosen
as $N_x \times N_y = 48 \times 24$ sites.

First let us choose $U=2.2$ such that AF order is completely
suppressed at zero field. Our calculation is performed at low
temperature. The spatial variation of DSC order parameter
$\Delta_i^{D}$ is plotted on a $24 \times 24$ lattice in Fig. 1(a)
with the vortex core situated at the center where the DSC order
parameter vanishes. By comparing it with the vortex structure of a
pure DSC, the size of the vortex core here is noticeably to be
enlarged. Fig. 1(b) displays the spatial variation of the induced
staggered magnetization of SDW order as defined by $M_{\bf
i}^{s}=(-1)^{i} \Delta^{SDW}_{\bf i}/U$. There the SDW order xists
both inside and outside the vortex cores , and exhibits isotropic
two-dimensional behavior with the period $8 a$ along both $x$ and
$y$ directions. Its magnitude reaches the maximum value at the
vortex core center. The DSC and SDW orders coexist throuhout the
whole sample. The appearance of the SDW order around the vortex
cores strongly affects the spatial profile of the local electron
density distribution, which can be represented by a weak CDW as
shown in Fig. 1(c). The remarkable enhancement of electron density
(or depletion of the hole density) is presented at the vortex core
center. The variation of the electron density outside the vortex
core shows weak oscillation. Different two-dimensional SDW and CDW
structures have also been obtained for a different set of band
parameters~\cite{chen01}.

Next we perform the calculation at very low temperature for
$U=2.4$, the obtained results are fundamentally different from
those for $U=2.2$, and they are presented in Fig. 2.  In Fig. 2(a)
we plot the spatial variation of DSC order parameter. It is clear
that $y$-axis oriented stripe like structures appear in
$\Delta_i^{D}$ with a weakly modulation period of $4 a$. The size
of vortex core is further enlarged and elongated along the $y$
axis than in Fig. 1(a). The SDW order and CDW order are displayed
in Fig. 2(b) and Fig. 2(c), respectively. The SDW order behaves
like almost uniform AF stripes oscillating with a wavelength of $8
a$. The vortex core is always pinned at one of the ridges of  AF
stripes where the AF order is stronger than those at other sites.
The spatial modulation of CDW order also exhibits
quasi-one-dimensional charge stripes behavior with a wavelength $4
a$, exactly half of that of the SDW along the $x$ direction. The
above numerical results are checked by three different set of
initial parameters $n_{{\bf i} \sigma}$ and $\Delta_{\bf i, j}$
and the iteration processes have been carried out for more than
500 steps to achieve the required accuracy. The above results for
finite B indicate that there exists a critical point $U_c \sim
2.25$ between $U=2.2$ and  $U=2.4$, such that isotropic
two-dimensional SDW and CDW may be induced when $U<U_c$ and they
become stripe like structures when $U>U_c$. As we shall show below
that the AF stripes and charge stripes obtained here could be very
relevant to experiments performed on the optimally doped BSCCO. It
is likely that some of the optimally doped and underdoped HTS
could be close to this critical region. At $B=0$ we found no
two-dimensional SDW and CDW regardless of the value of $U$.

Before comparing with experiments, we would like to point out that
the stripe phases oriented along $x$- and $y$- directions are
degenerate in energy. In order to compare with the experiments, we
shall assume that in certain domains of the sample the stripes are
further pinned by some defects which makes their orientation along
the $y$-direction more energetically favorable than those along
the $x$-direction. Setting the energy difference between the $x$-
oriented and $y$-oriented stripe phases to be ${\delta}E >0$ and
defining ${\eta} = exp(- {\delta}E/T)$ with $T$ as the
temperature, then the statistical probabilities for $y$-oriented
and $x$-oriented stripes to appear at temperature $T$ are
respectively $1/(1+ {\eta})$ and ${\eta}/(1+ {\eta})$. The
measured order parameter should be $O=[O(y) + \eta O(x)]/ (1+
\eta)$ with $O(x)$ $[O(y)]$ representing the order parameter for
one of the $x$- ($y$-) oriented DSC, AF and charge stripes. The
combined results are shown in Fig. 3 for ${\eta}=0.5$. The spatial
variation of the combined superconductivity order parameter is
presented in Fig. 3(a). Here the SDW (see Fig.3(b)) and CDW (see
Fig.3(c)) have anisotropic two dimensional structures and
respectively with periodicities $8 a$ and $4 a$, in good
agreements with the observations of Lake {\em et
al.}~\cite{Lake01} and Hoffman {\em et al.}~\cite{Hoff1}. Our
results also predict that as $T$ approaches zero or ${\eta}=0$, AF
and charge stripes should show up, and at higher $T$ or ${\eta}$
close to 1, the observed SDW and CDW should become more isotropic.
It is useful to point out that the way to explain these
experiments may not be unique. As suggested by Howald {\em et al.}
~\cite{coex}, if  $x$-oriented and $y$-oriented stripes are pinned
in two nearest neighboring domains, the proximity effect may cause
these two different oriented stripes permeating each other and
make their spatial distribution to look  more two- dimensional
like. So far we are not able to numerically simulate this
situation.

The features exhibited in Fig. 2 and Fig. 3 are very robust when
$U>U_c$. We believe that the optimally doped BSCCO sample should
have a $U>U_c$. While the STM experiments~\cite{Hoff1} were
performed at $B \simeq 7$  Tesla, our numerical results based upon
the $48 \times 24$-sites calculation corresponding to a magnetic
field $B \simeq 27$ Tesla. For a realistic comparison with
experiments, one needs to do a $92 \times 46$ -sites calculation,
that is beyond our current computing capability. In a weaker field
$B \simeq 7$ Tesla, we expect that the SDW and CDW structures with
periods $8a$ and $4a$ should still exist in the neighborhood of a
vortex core. Away form the vortex core, the situation should be
corresponding to $B=0$ case.

We have also done a numerical study for $B=0$ with $U=2.4$ and
$x=0.15$, and our calculation indicates that both the stripe phase
and the uniform DSC phase (no SDW and CDW) could show up depending
on the initial input parameters. Since $U=2.4$ is quite close to
$U_c$, the free energy difference between the stripe phase and the
uniform DSC phase is estimated to be very small, which suggests
that the experimental observed phase at finite temperature should
come from a superposition of these two configurations. This
consideration would dramatically reduce the amplitudes of the SDW
and CDW stripes measured by experiments. But when a magnetic field
B is applied, the stripe phase is the only solution regardless of
the initial parameters. This result implies that the SDW and CDW
orders are stabilized and enhanced near the vortex core within a
distance of several coherence lengths.  Away from the vortex core
they are somewhat suppressed. Of course, the stripe phase order at
$B=0$ could be stabilized by the presence of defects and it can
also be strengthened by a larger $U$. The enhancement of AF order
near the vortex core in a strong magnetic field is consistent with
neutron scattering experiments~\cite{Katano,Lake02}. Here The
manifestation of stripe phases in optimally doped sample seems to
be against common consensus. But our stripes are small
one-dimensional SDW and CDW modulations in a $d$-wave
superconducting background. This is different from what was
proposed originally by Emery and Kivelson~\cite{Emery} for
underdoped sample where the AF phases are insulators (no hole
regions) and charge stripes are conductors (rich hole regions).
There is no physical reason to forbid our stripe like modulations
appearing in optimum doped HTS. The observation of charge stripes
in very recent STM experiments~\cite{coex} at $B=0$ seems to
suggest that  stripe phases may indeed be present in optimally
doped BSCCO samples.

For the purpose to have a better understanding of the doping
effect, the underdoped case $(x=0.10)$ is examined. We found that
the spatial distribution of SDW and CDW still exhibits the
stripe-like behavior as shown in Fig. 2, and the periods of SDW
and CDW now change respectively to $12a$ and $6a$. We expect that
the periods could even become $16a$ and $8a$, when the doping
level is further reduced. These results are in qualitative
agreement with the observations of Lake {\em et al.}~\cite{Lake02}
where a magnetic field induced AF order was observed in underdoped
sample. A detailed study of the doping dependence will be reported
elsewhere.

Finally, we would like to point out that Zhu {\em et al.}
~\cite{Zhu2D}did almost the same calculation using slightly larger
$U=2.5$ and $x=0.16$, their results yield isotropic
two-dimensional SDW and CDW, which are very different from the AF
and charge stripes obtained by us. We have carefully checked their
calculations and found that if they increased the number of
iteration steps sufficiently to achieve higher accuracy, their
two-dimensional SDW and CDW checkerboard patterns would have
evolved into our stripe like structures as shown in Fig.2.

In summary, the stabilization and the enhancement of AF and charge
stripes in a $d$-wave superconductor near a vortex core are
numerically studied by a mean field Hamiltonian. We found the
wavelength of the AF stripes to be $8 a$ and that of charge
stripes to be $4 a$. Assuming that the degeneracy of the
$x$-oriented and $y$-oriented stripe phases is broken by some
defects, we show that the observed spatial variations of SDW and
CDW are anisotropic and two-dimensional, with wavelengths in good
agreement with experiments. We also would like to emphasize that
our self-consistent mean field BdG equation calculation tends to
overestimate the stability of the static AF order. In order to
partially overcome this deficiency, we choose the onsite Coulomb
interaction $U=2.4$ to be somewhat smaller than that for the
standard Hubbard model. The effect due to the dynamic SDW has not
been included in this study. Any attempt to do the present type of
calculation by including this effect is quite difficult and has to
be confined to much smaller  magnetic unit cell. This would make
the comparison with experiments difficult. As to whether a static
AF order could exist in optimally doped sample is still a subject
for debate. The coexistence of charge stripes with DSC at $B=0$
observed by very recent STM experiments ~\cite{coex} at optimal
doping may indicate that static AF stripes are also in presence.
For a pure DSC, the local density of states (LDOS) at the vortex
center is well known to have a broad peak around $E=0$
~\cite{Wang95}. However, the vanishing LDOS at $E=0$ near the
vortex core observed by STM experiments~\cite{Maggio,Pan00} for
YBCO and BSCCO  has been understood in terms of  the presence of
SDW in a $d$-wave superconductor~\cite{chen01}, an indirect
evidence of the existence of the static AF order in optimally
doped HTS samples. In view of all these and the favorable
comparison with experiments, The qualitative feature of our
results should still remain even in a more refined theory.

We wish to thank Prof. S. H. Pan and Dr. J. X. Zhu for useful
discussion.  This work was supported by a grant from the Robert A.
Welch Foundation and by the Texas Center for Superconductivity at
the University of Houston through the State of Texas.

\newpage
\begin{figure}\caption[*] {Spatial variations of the DSC order parameter $\Delta_{\bf i}^{D}$ (a),
staggered magnetization $M_{\bf i}^{s}$ (b), and electron density
$n_{\bf i}$ (c) in a $24 \times 24$ lattice. The size of a
magnetic unit cell is $48 \times 24$, corresponding to a magnetic
field $H= {\Phi_0}/(24 \times 24)$. The strength of the on-site
repulsion $U=2.2$ and the averaged electron density $\bar
{n}=0.85$.}\label{1}
\end{figure}

\begin{figure}\caption[*] {Spatial variations of the order parameters $\Delta_{\bf i}^{D}$ (a),
$M_{\bf i}^{s}$ (b), and $n_{\bf i}$ (c).  The strength of the
on-site repulsion $U=2.4$. The other parameter values are the same
as Fig. 1.}\label{2}
\end{figure}

\begin{figure}\caption[*] {Spatial profiles of the combined  $x$-
and $y$- oriented stripes with the order parameters $\Delta_{\bf
i}^{D}$ (a), $M_{\bf i}$ (b), and $n_{\bf i}$ (c). Only the $20
\times 20$ lattice with the vortex core at the center is plotted
in (c). The mixing factor is chosen as ${\eta}=0.5$. The other
parameter values are the same as Fig. 2.}\label{3}
\end{figure}


\begin{thebibliography}{19}
\vspace*{-1.0cm}
\bibitem{Katano} S. Katano {\em et al.}, Phys. Rev. B {\bf 62}, R14677 (2000).
\bibitem{Lake01} B. Lake {\em et al.}, Science {\bf 291}, 1759 (2001);
B. Lake {\em et al.}, cond-mat/0104026.
\bibitem{Lake02} B. Lake {\em et al.}, Nature {\bf 415}, 299 (2002).
\bibitem{Maggio} I. Maggio-Aprile {\em et al.}, Phys. Rev. Lett.
{\bf 75}, 2754 (1995).
\bibitem{Pan00} S. H. Pan {\em et al.}, Phys. Rev. Lett. {\bf 85}, 1536 (2000).
\bibitem{Hoff1} J. E. Hoffman {\em et al.},  Science {\bf 295}, 466 (2002).
\bibitem{NMR1} V. F. Mitrovic {\em et al.}, Nature {\bf 413}, 501 (2001).
\bibitem{Emery} V. J. Emery and S. A. Kivelson, Physica C {\bf 209}, 597
(1993); V. J. Emery, S. A. Kivelson, and J. M. Tranquada, Proc.
Natl. Acad. Sci. USA {\bf 96}, 8814 (1999);
\bibitem{Bale98} L. Balents, M. P. A. Fisher, and C. Nayak, Int. J. Mod.
Phys. B {\bf 12}, 1033 (1998).
\bibitem{Sach99} M. Vojta, and S. Sachdev, Phys. Rev. Lett. {\bf
83}, 3916 (1999).
\bibitem{Mart00} I. Martin, {\em et al.}, Int. J. Mod. Phys. B {\bf 14}, 3567 (2000).
\bibitem{Arovas97} D. P. Arovas {\em et al.}, Phys. Rev. Lett. {\bf 79},
2871 (1997); J.-P. Hu, and S. C. Zhang, cond-mat/0108273.
\bibitem{Machida} M. Ichioka, M. Takigawa, and K. Machida, J. Phys. Soc. Jpn. {\bf 70}, 33 (2001).
\bibitem{Demler} E. Demler, S. Sachdev, and Y. Zhang, Phys. Rev. Lett.
{\bf 87}, 067202 (2001); Y. Zhang, E. Demler, and S. Sachdev,
cond-mat/0112343.
\bibitem{Lee01} D. H. Lee, Phys. Rev. Lett. {\bf 88}, 227003 (2002).
\bibitem{Wang95} Y. Wang and A. H. MacDonald, Phys. Rev. B {\bf 52},
R3876 (1995).
\bibitem{Zhu01} Jian-Xin Zhu, and C. S. Ting, Phys. Rev. Lett. {\bf 87},
147002 (2001).
\bibitem{chen01} Yan Chen, and C. S. Ting, Phys. Rev. B {\bf 65}, R180513 (2002).
\bibitem{coex} C. Howald {\em et al.}, cond-mat/0201546.
\bibitem{Zhu2D} Jian-Xin Zhu, I. Martin, and A. V. Bishop,
cond-mat/0201519.
\end{thebibliography}
\end{document}